\documentclass{article}
\usepackage[T1]{fontenc}
\usepackage[cp850]{inputenc}
\usepackage[english]{babel}
\usepackage{euscript,amstext,amssymb,amsbsy,fancybox,epsf}

\def\beqn{\begin{eqnarray}}
\def\eeqn{\end{eqnarray}}
\def\barr{\begin{array}}
\def\earr{\end{array}}
\def\btab{\begin{tabular}}
\def\etab{\end{tabular}}
\def\bite{\begin{itemize}}
\def\eite{\end{itemize}}
\def\bcen{\begin{center}}
\def\ecen{\end{center}}

\begin{document}
\renewcommand{\thefootnote}{\fnsymbol{footnote}}

\begin{center}
{\Large Deeply Virtual Electroproduction\\
of Photons and Mesons.} 
\vskip 1. cm
{\bf M. Guidal$^a$, M. Vanderhaeghen$^b$}\\ 
{\it $^a$ IPN Orsay, F-91406 Orsay, France}\\
{\it $^b$ University Mainz, D-55099 Mainz, Germany}
\end{center}

Much of the internal structure of the nucleon has been revealed
during the last two decades through the \underline{inclusive}
scattering of high energy leptons on the nucleon in the
Bjorken -or "Deep Inelastic Scattering" (DIS)- regime 
($Q^2,\nu\gg$ and $x_B=\frac{Q^2}{2M\nu}$ finite). 
Simple theoretical 
interpretations of the experimental results and quantitative
conclusions can be reached in the framework of QCD, when 
one sums over all the possible hadronic final states. For 
instance, {\it unpolarized} DIS
brought us evidence of the quark and gluon substructure 
of the nucleon, quarks carrying about 45\% of the nucleon
momentum. Furthermore, {\it polarized} DIS revealed that about 
25\% of the spin of the nucleon is carried by the quarks' 
intrinsic spin.\\
Now, with the advent of the new generation of high-energy,
high-luminosity lepton accelerators combined with large
acceptance spectrometers, a wide variety of 
\underline{exclusive} processes in the Bjorken regime can 
be envisaged to become accessible experimentally. Until
recently, no sound theoretical formalism could really allow to
interpret in a unified way such processes. It now appears
that such a coherent description is under way through the
formalism of new generalized parton distributions, the so-called
'Off-Forward Parton Distributions' (OFPD's). It has been
shown that these distributions, which parametrize the structure 
of the nucleon, allow to describe, in leading order perturbative
QCD (PQCD), various exclusive processes such as, in particular,
Virtual Compton Scattering (\cite{Ji97}\cite{Rady})
and (longitudinal) vector and pseudo-scalar meson electroproduction 
\cite{Collins97}. Maybe most importantly, Ji \cite{Ji97} showed 
that the second moment of these OFPD's gives access to the sum
of the quark spin and the quark orbital angular momentum to the nucleon
spin, which may shed light on the "spin-puzzle".\\
In this paper, after a brief summary of the properties of the OFPD's, 
we give some examples of what could be 
the experimental opportunities to access the OFPD's at the current 
high-energy lepton facilities~: JLab ($E_e\geq$ 6 GeV), HERMES 
($E_e$=27 GeV) and COMPASS ($E_\mu$=200 GeV).\\

\begin{figure}[ht]
\epsfxsize=11 cm
\epsfysize=6. cm
\centerline{\epsffile{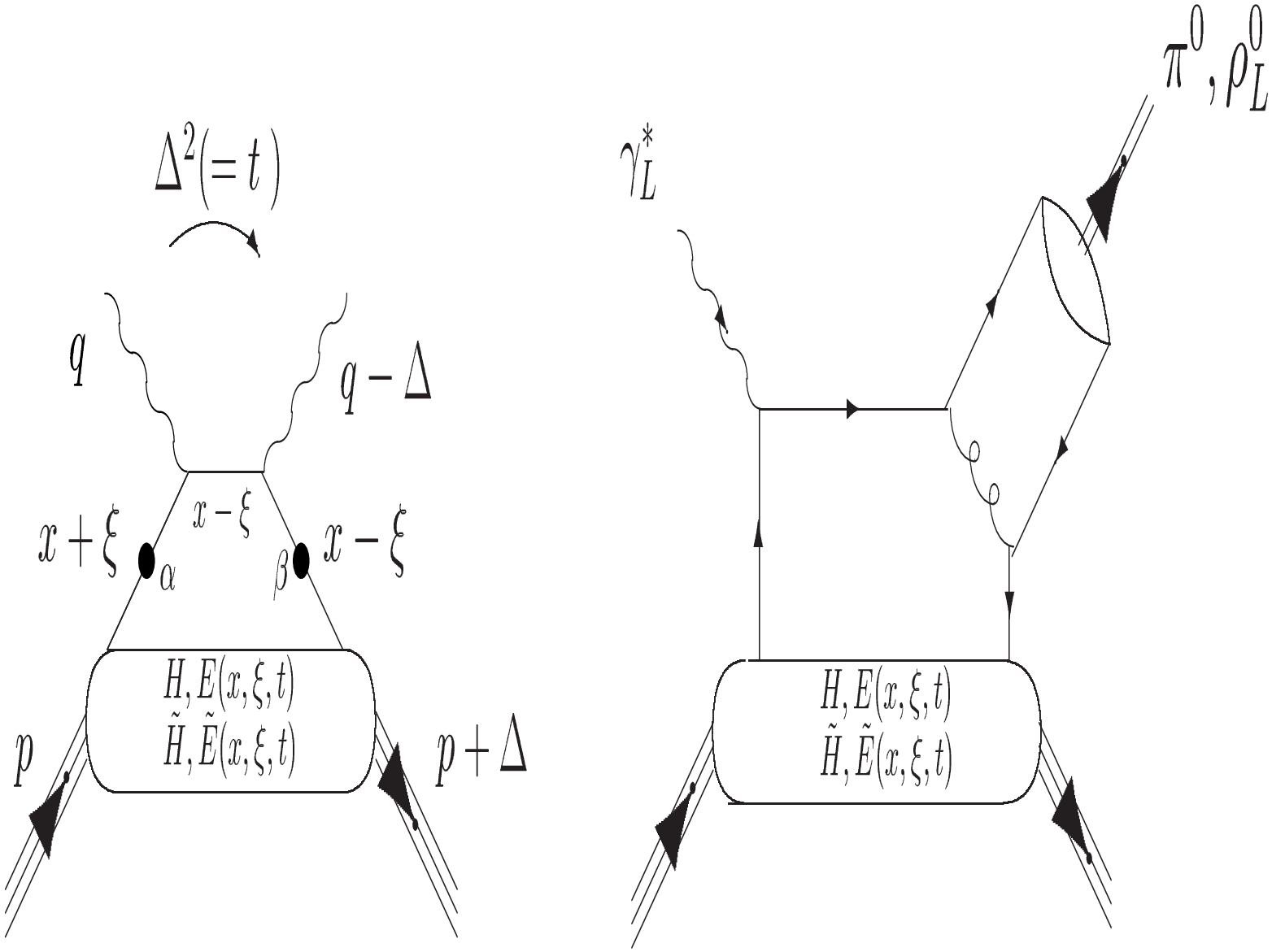}}
\vspace{-1.5cm}
\caption[]{"Handbag" diagrams~: a) for DVCS (left) and b) for meson 
production (right).}
\label{fig:handbags}
\end{figure}

Recently, Ji \cite{Ji97} and Radyushkin \cite{Rady}
have shown that the leading order PQCD DVCS amplitude in
the forward direction can be factorized in a hard scattering part 
(exactly calculable in PQCD) and a nonperturbative nucleon 
structure part as is illustrated in Fig.(\ref{fig:handbags}-a). 
In these so-called ``handbag" diagrams of Fig.(\ref{fig:handbags}), 
the lower blob which represents the structure of the nucleon 
can be parametrized, at leading order PQCD, in 
terms of 4 generalized structure functions, the OFPD's. 
These are defined as $H, \tilde H, E, \tilde E$, and  
depend upon three kinematical invariants~: $x$, $\xi$, $t$. 
$H$ and $E$ are spin independent  
and $\tilde H$ and $\tilde E$ are spin dependent.

The OFPD's $H$ and $\tilde H$ are  
actually a generalization of the parton distributions 
measured in deep inelastic scattering. Indeed, in the forward 
direction, $H$ reduces to 
the quark distribution and $\tilde H$ to the 
quark helicity distribution measured in deep inelastic scattering. 
Furthermore, at finite momentum transfer, there are
model independent sum rules which relate 
the first moments of these OFPD's to the elastic form factors.
The OFPD's reflect the structure of the nucleon independently of
the reaction which probes the  nucleon. They can also be accessed 
through the hard exclusive electroproduction of mesons 
-$\pi^0$, $\rho^0$, $\omega$, $\phi$,...- 
(see Fig.(~\ref{fig:handbags}-b)) for which a QCD factorization
proof was given recently \cite{Collins97}. According to 
Ref.\cite{Collins97}, the factorization applies when the virtual photon is 
longitudinally polarized because in this case, the end-point contributions
in the meson wave function are power suppressed. 
It was also shown in Ref.\cite{Collins97} that the cross section for 
a transversely polarized photon is suppressed by 1/$Q^2$ compared to 
a longitudinally polarized photon. 
Because the transition at the  upper vertices of Fig.(~\ref{fig:handbags}-b) 
will be dominantly helicity conserving 
at high energy and in the forward direction, 
this means that the vector meson will also be predominantly longitudinally 
polarized (notation $\rho^0_L, \omega_L, \phi_L$) for a longitudinal 
photon. By identifying then the polarization of the vector meson 
through its decay angular distribution, one can obtain 
the longitudinal part of the electroproduction cross sections.\\ 
It was also shown in \cite{Collins97} that leading order PQCD predicts
that the vector meson channels ($\rho^0_L$, $\omega_L$, $\phi_L$)
are sensitive only to the unpolarized OFPD's ($H$ and $E$) whereas
the pseudo-scalar channels ($\pi^0, \eta,...$) are sensitive only to the 
polarized OFPD's ($\tilde{H}$ and $\tilde{E}$). In comparison to meson electroproduction, 
we recall that DVCS depends at the same time on {\it both} the polarized and 
unpolarized OFPD's.\\
For a first exploratory approach, we will now show that the
meson channels hold the best promises due to the relatively high
cross-sections. First estimates for
the  $\pi^0$, $\rho^0_L$ cross sections were given in 
Refs.\cite{marcprl} \cite{vcsrev} besides the $\gamma$-channel 
using an educated guess for the OFPD's, 
which consists of a product of elastic form factors by 
quark distributions measured in DIS. This ansatz satisfies the first sum rules and 
the corresponding distributions obviously reduce to the quark
distributions from DIS in the forward direction.\par

\begin{figure}[h]
\vspace{-2cm}
\hspace{.5cm}
\epsfxsize=12. cm
\epsfysize=10. cm
\centerline{\epsffile{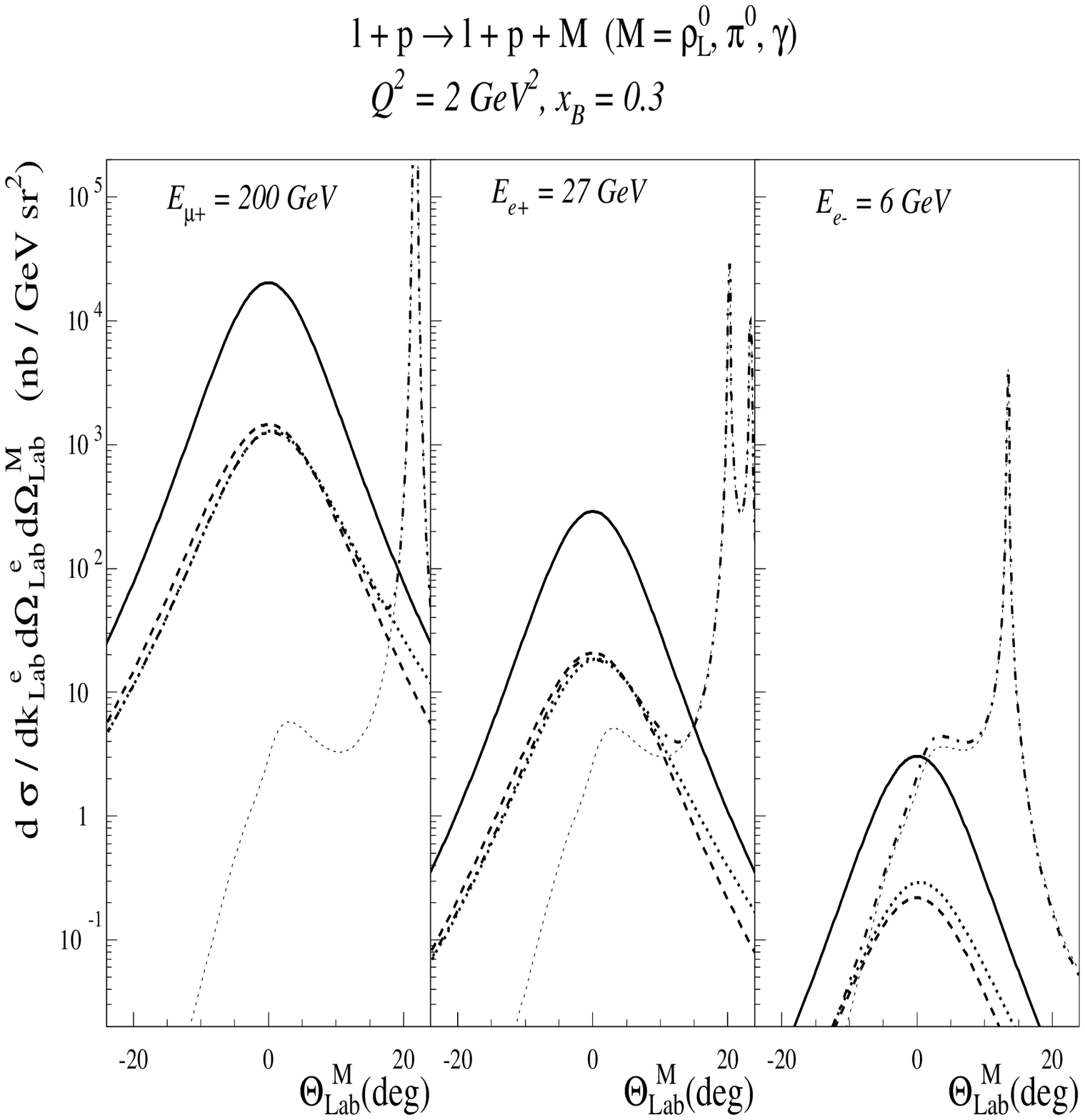}}
\caption[]{\small Comparison between $\rho^0_L$ (full lines), 
$\pi^0$ (dashed lines), DVCS (dotted lines), BH (thin dotted lines) 
and total $\gamma$ (dashed-dotted lines) 
leptoproduction in-plane cross sections at $Q^2$ = 2 GeV$^2$, $x_B$ = 0.3 
and for different beam energies : 
$E_{\mu^+}$ = 200 GeV (COMPASS), $E_{e^+}$ = 27 GeV (HERMES), 
$E_{e^-}$ = 6 GeV (CEBAF).}
\label{fig:diffkin}
\end{figure}

We compare in Fig.(\ref{fig:diffkin}), 
the $\rho^0_L$, $\pi^0$ and $\gamma$ cross sections as function of 
the beam energy at a fixed $Q^2$ = 2 GeV$^2$ and $x_B$ = 0.3. 
It is clear on this picture that the $\rho$ channel is very favorable.
Its cross section is the highest because it depends on the 
unpolarized OFPD's ($H$ and $E$). The $\omega_L$ channel has 
a cross section that is substantially higher than the ratio 
$\sigma_\omega$/$\sigma_\rho=\frac{1}{9}$ predicted by 
the diffractive mechanism and this is essentially due to the 
quark exchange mechanism (QEM). 
The $\omega_L$ and $\rho^0_L$ channels probe 
different combination of the $u$ and $d$ OFPD's and a 
measurement of both therefore allows 
to separate these $u$ and $d$-quark unpolarized OFPD's.
The $\pi^0$ channel depends on 
the polarized OFPD's ($\tilde{H}$ and $\tilde{E}$) 
and therefore the PQCD QEM mechanism gives a lower cross section. The DVCS is proportional
to {\it both} the polarized and the unpolarized OFPD's as was already 
mentioned but it has an extra $\alpha_{em}$ coupling (due to the final 
state photon) which reduces the cross section. (By comparison, the meson 
final states go through the exchange of a gluon and therefore has a $\alpha_S$
coupling). Furthermore, at JLab energies, the DVCS suffers from
the competing process which leads to the same final state, the Bethe-Heitler
process. This extra ``parasite" mechanism is dominant at 6 GeV and renders
the extraction from the cross section of the DVCS process very
difficult. This ``parasite'' process is absent in the case of meson
electroproduction. 
Going up in energy, the increasing virtual photon flux factor boosts the 
$\rho^0_L$, $\pi^0$ leptoproduction cross sections and the DVCS part of 
the $\gamma$ leptoproduction cross section.
For the $\gamma$ electroproduction cross section the BH process 
is hardly influenced by the beam 
energy and therefore overwhelms the DVCS cross section at low beam energies. 
For a study of the OFPD's such a figure seems to favor high-energy 
experimental facilities such as COMPASS. However, it should be clearly
kept in mind that the actual count rates will be weighted by the 
luminosity. So, in spite of the relatively "low" energy of the JLab
incident beam, the higher luminosity and the better resolution that
one can reach with the JLab large acceptance CLAS detector will allow equivalent 
count rates to the other two facilities in the roughly same 
equivalent range (but in a shorter period).
 


\begin{figure}[h]
\epsfxsize=11 cm
\epsfysize=9 cm
\centerline{\epsffile{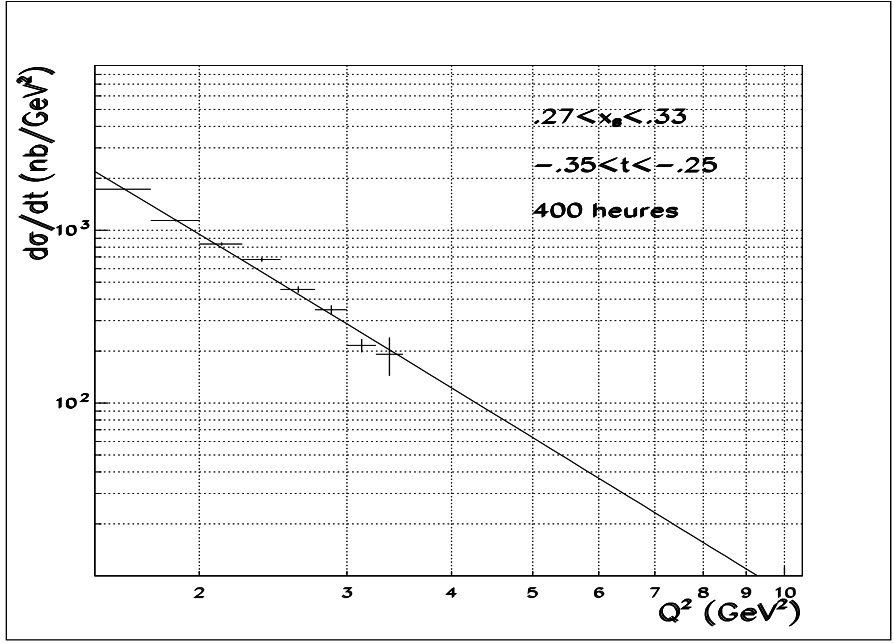}}
\caption[]{Error estimate (statistical only)
on the scaling behavior for the $\rho^0_L$ channel for $t=-.3$ GeV$^2$
and $x_B=.3$.}
\label{fig:scalexp}
\end{figure}

Before considering the extraction of the OFPD's from the data,
it is mandatory to first demonstrate that the scaling regime has been 
reached. In leading order PQCD,
the DVCS transverse cross section $\frac{d\sigma_T}{dt}$ is predicted
to behave as $\frac{1}{Q^4}$ whereas the mesons'
longitudinal cross sections will obey a $\frac{1}{Q^6}$
scaling (due to the "extra" gluon exchange for the mesons,
see Fig.~\ref{fig:handbags}). Recently, an experiment
\cite{propo98107} has been approved at JLab where it is proposed 
to investigate this scaling behavior.
Figure~\ref{fig:scalexp} shows the estimated lever arm 
reachable at JLab with 400 hours of beam time in the CLAS detector
for the $\rho^0$ channel.
With a maximum $Q^2$ of $\approx$ 3.5 GeV$^2$ (for $x_B$ around .3),
the cross section can be measured over about a decade. This should 
provide a sufficient lever arm to test the scaling prediction
and test at what value of $Q^2$ this $\frac{1}{Q^6}$ scaling 
behavior sets in. With a JLab 8 GeV incident energy, the lever
arm extends to 4.5 GeV$^2$.\\

In conclusion, we believe that a broad new physics program, 
i.e. the study of \underline{exclusive} reactions at large $Q^2$ 
in the valence region (where the quark exchange mechanism dominates), opens up.
By "constraining" the final state of the DIS reaction, instead of
summing over all final states, one accesses some more fundamental structure
functions of the nucleon, i.e. the OFPD's. These functions provide
a unifying link between a whole class of various reactions 
(elastic and inelastic) and fundamental quantities as diverse as
form factors, parton distributions, etc...

\end{document}